\documentclass{article}
\usepackage{graphics}
\newcommand{\ket}[1]{{|#1\rangle}}

\title{Fast versions of Shor's quantum factoring algorithm}

\author{Christof Zalka\footnote{Supported by
Schweizerischer Nationalfonds and LANL} \\ zalka@t6-serv.lanl.gov}

\begin{document}
\maketitle

\begin{abstract}
We present fast and highly parallelized versions of Shor's
algorithm. With a sizable quantum computer it would then be possible
to factor numbers with millions of digits. The main algorithm
presented here uses FFT-based fast integer multiplication. The quick
reader can just read the introduction and the ``Results'' section.
\end{abstract}
\newpage
\tableofcontents

\section{Introduction}

\subsection{Motivation}

The algorithms presented here are useful to factor very large numbers,
that is, thousands to millions of digits. Quantum computers which
could do that will at best be available in several decades. Still I
think it is worth it to investigate already now what such machines
could do. Also it shows that messages encrypted even with very large
RSA\footnote{RSA is a widely used public-key cryptosystem whos
security relies on the difficulty to factor large numbers} keys could
possibly be decrypted in several decades. 

\subsection{Summary}

\subsection{assumptions on future quantum computer architectures}

We don't know the architecture and power of future (sizable\footnote{I
think one shouldn't call a 2 or 3 quantum bit system a quantum
computer}) quantum computers. It may therefore seem premature to
optimize a quantum algorithm and I admit that my results should be
considered as a rough guide rather than precise predictions. I make
plausible assumptions on the architecture of future quantum computers
which I will try to justify below. My main assumptions are that QC's
will anyways be parallel, that qubits will be expensive, that QC's
will have a high ``connectivity'' between its qubits, that QC''s may
be slow and that fault tolerant techniques (quantum error correcting
codes) will be used. That means that I'm looking for highly
parallelizable algorithms which nevertheless don't use much more space
(qubits) than simpler algorithms. If there are plenty of qubits,
additional parallelization schemes could be used with a relatively bad
space-time tradeoff of the form $T_p \sim 1/S$. Because present
propositions for fault tolerant techniques \cite{shor2} are especially
slow for Toffoli gates, I will only count those for my performance
analyses.

\subsection{First algorithm: standard but with parallelized addition}

\subsubsection{improved standard algorithm with $S=3 L$ and $T=12 L^3$}

First I give an improved version of the standard algorithm
\cite{miquel} which uses only $3 L$ qubits and some $12 L^3$ Toffoli
gates, where $L$ is the number of bits of the number to be
factored. The ideas that lead to these improvements are due to several
people. My contribution is the observation that it is enough to
compute the modular exponentiation correctly for most but not all
input values, as it should still be possible to extract the period of
this function is a few runs of the quantum computer. This allows
substantial simplifications of the algorithm. The idea is used
throughout this paper.

\subsubsection{parallelizing addition}

The standard way to compute the modular exponentiation \cite{miquel}
is to decompose it into many modular multiplications which again are
decomposed into additions. Usually the addition has to be done bit
after bit. I propose a way to parallelize it such that the execution
time essentially becomes a constant for large numbers. The space
requirements of the algorithm increase from $S=3 L$ to $S=5 L$ qubits.

\subsection{Second algorithm: using FFT-based fast integer multiplication}

Here we directly ``attack'' the modular multiplication by using the
fast Fourier transform (FFT) -based multiplication technique which led
to the famous Sch\"onhage-Strassen algorithm. 

Note that the FFT here has nothing to do with the quantum Fourier
transform. The latter Fourier-transforms the amplitudes of a quantum
register. The former is a classical operation, even though it is
applied to a superposition and thus is computed in ``quantum
parallelism''. Applied to a ``classical'' basis state it computes the
Fourier transform of the values which are represented in binary in
several registers.

FFT-based fast integer multiplication is rather complicated, it
consists of several ``subroutines'' which I have figured out how to do
reversibly. Also the Fourier transform employed is not the usual one
over the complex numbers, but over the finite ring of integers modulo
some fixed integer. FFT-multiply reduces the multiplication of two big
numbers to many smaller multiplications. I will use the same technique
again to compute these smaller multiplications, thus I propose a
2-level FFT-multiply.

I have investigated numerous other versions, like 1-level FFT-multiply
with parallelized addition or the $O(n^{\log_2 3})$ Karatsuba-Ofman
algorithm or FFT-multiply using a modulus which is not of the form
$M=2^n+1$ as is usually used. These algorithms seemed not to perform well
enough either on space (Karatsuba-Ofman) or on time and I'm not
discussing all these possibilities.

\section{Assumptions on future quantum computer architectures}

As mentioned above, my main assumptions are that quantum computers
will be parallel, that qubits will be expensive and that communication
across the QC will be fast (contrary e.g. to a cellular automaton).

The parallelism assumption comes from the following observations:
Qubits have to be controlled by exterior field which are again
controlled by a classical computer. Probably every qubit or few qubits
will have its own independent (classical) control unit. This also
explains why I think that qubits will be expensive, thus we want to
use as few of them as possible. Also if decoherence is strong we will
have a high rate of memory errors acting on resting qubits. This will
make necessary periodic error recovery operations (quantum error
correction) also on resting qubits, thus the QC must anyways be
capable of a high degree of parallelism. Note that in the case where
memory errors dominate over gate inaccuracy errors we actually loose
nothing by running the computation in parallel, as it will not
increase the error rate by much.

Here I have of course assumed that large quantum computations will
need fault tolerant techniques (see e.g. Shor \cite{shor2}). Thus all
``computational'' qubits will be encoded in several physical qubits
and operations (gates) will be done on the encoded qubits, thus
without decoding. Presently proposed schemes (Shor \cite{shor2},
Gottesman \cite{gottesman}) are much slower on Toffoli gates than on
the other used gates, which is why I propose to only count the Toffoli
gates when assessing the execution time. 

I assume that better fault tolerant schemes will be found, e.g. ones
using less space than the very space-consuming concatenation technique
(see Knill et al. \cite{knill} or Aharonov et al \cite{aharonov}).
Also I think that faster ways of implementing the Toffoli gate fault
tolerantly will be found.

Actually there is an argument (Manny Knill, private communication)
which shows that it is not possible to implement all gates of a set of
universal gates transversally, that is bitwise between the individual
qubits of the encoded computational qubits. That is because otherwise
errors that don't lead out of the code space could happen, which
couldn't be detected. Present schemes allow the transversal
implementation of CNOT, NOT and Hadamard transform. Therefore in these
schemes it is not possible to implement the Toffoli gate
transversally. I now propose to look for a scheme where all the
``classical'' gates Toffoli, CNOT and NOT can be implemented
transversally, but not non-classical gates like the Hadamard. This
would help a lot for Shor's algorithm (and probably also for
applications of Grover's algorithm) as there except for a few gates at
the eginning and at the end, we use only such ``classical'' gates.

Why do I think that QC's may be much slower than todays conventional
computers? I think that 2-bit gates (or multi-bit gates) will be slow,
as it is not easy to keep qubits from interacting with each other
sometimes and then (with exterior fields) to make them interact
strongly. Actually this is essentially true for all present quantum
computer hardware proposals.

The assumption that fast quantum communication will be possible
between (relatively) distant parts of a QC is not well founded, rather
I think it will simply be a necessity for quantum computations. It
should be possible for QC's which either use photons to represent
qubits or where a coupling to photons is possible. Then connections
could simply be optical, possibly simply with optical fibers. In the
ion trap scheme it is clear that a large QC could hardly consist of a
single ion trap. Cirac and Zoller have thus extended their proposal to
couple ion-trap qubits to photons. 

The degree of connectivity may still be less than desired. When
interpreting my results for computation time, one must keep in mind
that things might actually be worse for a realistic quantum computer.

A general assumption that I make is that measurements of qubits will
not be too hard, as otherwise fault tolerance would be much harder to
achieve.

I also assume that classical computation will be much cheaper (and
probably also faster) than quantum computation. Therefore where
possible I (pre-) compute things classically. (Of course this makes
sense only up to a certain point.)

\section{The standard algorithm}

The most part of Shor's quantum factoring algorithm consists of
computing the modular exponentiation. This can be seen as a classical
computation, as it transforms computational basis states into
computational basis states. Of course this transformation is applied
``in quantum parallelism'' to a large superposition of such
states. Still we can think of it as being applied to just one basis
state. Then it differs from conventional computation only in that it
must be reversible.

So for Shor's algorithm we have to compute:

\begin{equation}
x \to x,a^x ~mod~ N
\end{equation}

Any conventional algorithm can also be run on a reversible machine,
except that then a lot of ``garbage'' bits will be produced. If we are
content with leaving around the input (here $x$), there is a general
procedure to get rid of the garbage. Say we want to compute $f(x)$,
but necessarily also produce garbage $g(x)$. Then we can do the
following to get rid of it:

\begin{equation} \label{rev1}
x \to x,g(x),f(x) \to x,g(x),f(x),f(x) \to x,0,0,f(x)
\end{equation}

In the $2^{nd}$ step we copy $f(x)$ to an auxiliary register,
initialized to 0. This can simply be done by bitwise CNOT. The last
step is the time reverse of the first, thus it ``uncomputes'' $g(x)$
and the first copy of $f(x)$. In general the work space will have a
size about the number of operations of the computation. Fortunately we
can do much better for the modular exponentiation, as we will see
later.

How can one compute efficiently a modular exponentiation with a large
exponent? The method is:

\begin{equation}
a^x ~mod~ N= a^{\sum x_i 2^i} ~mod~ N = 
\Pi_i \left( a^{x_i 2^i} ~mod~ N \right) ~mod~ N
\end{equation}

Where the $x_i$ are the bits of the binary expansion of $x$.  The
numbers $a^{(2^i)} ~mod~ N$ can be calculated by repeated
squaring. The modular exponentiation is then computed by modular
multiplication of a subset of these numbers. We will have a ``running
product'' $p$ which we will modularly multiply with the next candidate
factor $a^{(2^i)} ~mod~ N$ if $x_i$ is 1. Reversibly we will do:

\begin{equation}
p \to p, p \cdot A ~mod~ N \to 0, p \cdot A ~mod~ N
\end{equation}

The $2^{nd}$ step is possible because modular multiplication with $A$
is a 1 to 1 function and furthermore because we know how to
efficiently compute its inverse. The general scheme in reversible
computation for such a situation is:

\begin{equation} \label{rev2}
x \to x,f(x) \to 0,f(x)
\end{equation}

For the $2^{nd}$ step imagine that we applied the inverse of $f$ to
$f(x),0$, thus obtaining $f(x),x$. So the $2^{nd}$ step above is
essentially the time reverse of this. For modular multiplication the
inverse function is simply the modular multiplication with the inverse
of $A$ modulo N. $A$ has such an inverse because it is relatively prim
with $N$. This is because we assume that the constant $a$ is
relatively prime with $N$. Then $A^{-1} ~mod~ N$ can easily be
precomputed classically using Euclid's algorithm for finding the least
common divisor of 2 numbers.

Let's now concentrate on how to compute 

\begin{equation}
p \to p, p \cdot A ~mod~ N
\end{equation}

we can make a further simplification:

\begin{equation}
p \cdot A ~mod~ N = \sum p_i 2^i \cdot A ~mod~ N 
=\left(\sum p_i (2^i A ~mod~ N) \right) ~mod~ N
\end{equation}

Again the numbers $2^i A ~mod~ N$ can be precomputed classically. So
now we have reduced modular multiplication to the addition of a set of
numbers of the same length as $N$.

For this modular addition there are 2 possibilities: either we first
add all numbers not modularly and then compute the modulus, or we have a
``running sum'' $s$ and every time we add a new number to it we
compute the modulus. At least for conventional computation the
$1^{st}$ possibility is preferable: Say $L$ is the number of bits of
$N$, and thus also the length and the number of the summands. The
total sum will then be some $\log_2 L$ bits longer than $L$. To
compute the modulus of this number will take some $\log_2 L$ steps,
whereas we have to do some $L$ steps if we compute the modulus at each
addition. For reversible computation the problem is that computing the
modulus of the total sum is not 1 to 1 whereas modular addition (of a
fixed ``classical'' number) is. Fist I will describe the modular
addition technique and later I will show that the more efficient
method is also possible in reversible computation.

Let's first see how we can make a (non-modular) addition of a fixed
(``classical'') number to a quantum register. Actually this can be
done directly without leaving the input around. So we want to do

\begin{equation}
s \to s+B
\end{equation}

Addition of course is done binary place by binary place, starting with
the least significant bit. For every binary place we will also have to
compute a carry bit. In reversible computation we will need a whole
auxiliary register to temporarily hold these carry bits before they
get ``uncomputed''. Thus we will do

\begin{equation}
s \to c,s+B \to s+B
\end{equation}

Say $c_i$ is the carry bit that has been calculated from the place
number $i-1$, thus we will want to add it to the bit $s_i$. Then for
every place we have to do the following operations:

\begin{equation}
c_{i+1} = (s_i+B_i+c_i \ge 2) \qquad s_i \to s_i \oplus B_i \oplus c_i
\end{equation}

Here the parenthesis means a logical expression and $\oplus$ means XOR
which is the same as addition modulo 2. To show how to uncompute the
carries later, it is preferable to first compute the new $s_i$
and then compute the carry $c_{i+1}$:

\begin{eqnarray}
&& s_i \to s_i \oplus B_i \oplus c_i \\
&& \mbox{if}~ B_i=0 :~~ c_{i+1} 
= c_i ~\mbox{AND}~ \bar{s_i} \qquad 
\mbox{if}~ B_i=1 :~~ c_{i+1} = c_i ~\mbox{OR}~ \bar{s_i} 
\end{eqnarray}

The AND and OR can be realized by using a Toffoli gate
(CCNOT). Thus for either value of $B_i$ we need a Toffoli gate per
binary place and later another Toffoli gate to uncompute the carry.

Actually we want to add only if some conditional quantum bit is 1. We
could now make every gate conditional on this bit. Thus a NOT would
become a CNOT, a CNOT a CCNOT and so on. This would increase the cost
of the computation a lot, as much more Toffoli gates would be used,
e.g. a CCCNOT needs 3 Toffoli gates. Therefore it is preferable to do
as little as possible conditional on the ``enable-qubit'' (which
decides whether we should add or not). For addition I propose the
following: when uncomputing the carries, we also uncompute the $s_i$'s
conditional on the negation of the ``enable-qubit''. The computation
of the $s_i$'s usually needs only CNOT's and no Toffolis, so by making
this operation conditional on the enable qubit we will avoid costly
things like CCCNOT.

The following quantum network shows these operations for 1 binary place:

\newcommand{\NOT}{\begin{picture}(10,10)
    \put(0,0){\circle{10}} \put(0,-5){\line(0,1){10}}  
    \put(-5,0){\line(1,0){10}} \end{picture}}
\newcommand{\C}{\circle*{5}}

\begin{picture}(300,120)

\put(50,100){$B_i=0$} 
\thinlines \multiput(20,0)(0,25){4}{\line(1,0){120}} \thicklines
\put(0,75){$\bar{e}$} \put(0,50){$s_i$} \put(0,25){$c_i$} \put(0,0){$c_{i+1}$}

\multiput(50,50)(0,5){5}{\line(0,1){2}} \put(50,25){\line(0,1){25}}
\put(50,25){\C} \put(50,75){\C} \put(50,50){\NOT}

\put(110,0){\line(0,1){50}}
\put(110,0){\NOT}
\put(110,50){\C}
\put(110,25){\C}
\put(95,50){\NOT}
\put(125,50){\NOT}

\put(250,100){$B_i=1$} 
\thinlines \multiput(220,0)(0,25){4}{\line(1,0){120}} \thicklines
\put(200,75){$\bar{e}$} \put(200,50){$s_i$} \put(200,25){$c_i$} 
\put(200,0){$c_{i+1}$}

\multiput(240,50)(0,5){5}{\line(0,1){3}}
\put(240,75){\C} \put(240,50){\NOT}

\multiput(260,50)(0,5){5}{\line(0,1){3}} \put(260,25){\line(0,1){25}}
\put(260,25){\C} \put(260,75){\C} \put(260,50){\NOT}

\put(295,25){\NOT} 
\put(310,0){\line(0,1){50}}
\put(310,0){\NOT} \put(310,50){\C} \put(310,25){\C}
\put(325,0){\NOT}

\end{picture} \\

Here the dashed lines mean that only when we run this algorithm
backwards, should these gates also depend on the negated enable bit
$\bar{e}$.

Now let's look at modular addition

\begin{equation}
s \to s' = s+B ~mod~ N \qquad \mbox{where} \quad 0 \le s,B < N
\end{equation}

Note that $s'=s+B$ if $s < N-B$ and $s'=s+B-N$ if $s \ge N-B$. To do this,
we first compute the condition bit $(s \ge N-B)$, and then, depending
on it, we add $B$, resp. $B-N$. How can we uncompute the condition
bit? To do this we have to be able to compute it from $s'$. It is
easy to see that it is simply $(s' < B)$. Formally:

\begin{equation}
s \to (s \ge N-B),s \to (s \ge N-B),s' \to 
\underbrace{(s \ge N-B) \oplus (s' < B)}_{=0}
\end{equation}

Of course $B-N$ is negative, so we have to add the complement $2^n+B-N$,
where $2^n$ is the smallest power of 2 which is larger or equal to
$N$. After this addition we then simply have to flip the bit with
place value $2^n$ in the result.

Let's look at the second step in the above equation where the addition
is done. Thus at the $i^{th}$ binary place the classical bit we want
to add is either $B_i$ or $(2^n+B-N)_i$. If these two bits are equal
we can use the simple addition network described above. For the other
2 cases where $B_i \not = (2^n+B-N)_i$ the network is more
complicated, even though I have found a way to do it with the same
number of Toffoli gates. I describe below only the case $B_i=0$ and
$(2^n+B-N)_i=1$ as the remaining case is very similar. So:

\begin{picture}(300,145)

\put(50,125){$B_i=0 \qquad (2^n+B-N)_i=1$} 
\thinlines \multiput(70,0)(0,25){5}{\line(1,0){250}} \thicklines
\put(0,100){$\bar{e}$} \put(0,75){$(s \ge N-B)$} \put(0,50){$s_i$}
\put(0,25){$c_i$} \put(0,0){$c_{i+1}$}

\put(90,0){
\put(0,25){\line(0,1){50}} \put(0,75){\C} \put(0,25){\NOT}}

\put(120,0){
\multiput(0,50)(0,5){10}{\line(0,1){3}} \put(0,25){\line(0,1){25}}
\put(0,100){\C} \put(0,25){\C} \put(0,50){\NOT}}

\put(180,0){
\put(0,50){\line(0,1){25}} \put(0,75){\C} \put(0,50){\NOT}}

\put(200,50){\NOT}
\put(220,0){\line(0,1){50} \put(0,50){\C} \put(0,25){\C} \put(0,0){\NOT}}
\put(240,50){\NOT}

\put(260,0){
\put(0,50){\line(0,1){25}} \put(0,75){\C} \put(0,50){\NOT}}

\put(290,0){
\put(0,0){\line(0,1){75}} \put(0,75){\C} \put(0,0){\NOT}}

\end{picture} \\

The dashed line should be left away when we first compute the sum but
should be solid (thus here forming a Toffoli gate) when we uncompute
the carries (and possibly also the sum). To see that this network does
what it should, consider separately the cases $a_i=0$ and $a_i=1$. In
the first case the carry $c_{i+1}$ should only be set if $s_i=c_i=1$
and in the other case it should always be set except when
$s_i=c_i=0$. It is easy to check that this works out as it should.

Let's now look at the computation of the comparison (qu)bits. A
comparison seems to cost about as much as an addition. Say we start
the comparison from the most significant bits downwards. For 2 random
numbers we will usually see after a few bits which number is larger,
only if the two numbers are equal or differ only in the least
significant bit, have we to go through all bits. Of course in quantum
parallelism we can't do that, as there will usually always be some
fraction of the superposition (of computational basis states) which
are still equal for all bits we have looked at. But, I argue, that
once this fraction is small enough, we don't really have to care about
it.

More generally I say that if we compute the modular exponentiation
wrongly for some small fraction of input values $x$, it doesn't
matter much. Say

\begin{equation}
x \to x, a^x ~mod~ N + E(x) \qquad \mbox{where} \quad E(x) =0 \quad 
\mbox{for most}~ x
\end{equation}

Say the fraction of the input values $x$ in the superposition for
which $E(x) \not =0$ is equal to $\epsilon \ll 1$. The scalar product
of this state and the intended state then is $1-\epsilon$. It is clear
that there will be little change in the distribution of output values
observed at the end of the quantum computation. Anyways it is clear
that a quantum computer will work imperfectly and the error level we
expect will not be small. This may well still be true with error
correction techniques, as these techniques are very expensive
(especially in space) so that we will only use them as much as
necessary. Therefore the classical post-processing will anyways have
to take such errors into account. 

So I propose to simplify the modular exponentiation computation by
allowing ``algorithmic'' (or ``deterministic'') errors. In particular
here I propose to only compare some of the most significant bits of
two numbers to be compared. If these bits are equal for the two
numbers, we e.g. say that the 1. number is the larger. I make here the
plausible assumption that for estimating the error rate I can think of
the numbers as uniformly distributed random numbers. Mathematically
(and therefore very cautiously) inclined people \footnote{Manny Knill,
private objection} have questioned the validity of this
assumption. Here I simply assume that it is true, but note that one
could heuristically test it by running the simplified modular
exponentiation algorithm on a conventional computer for many inputs
and check the error rate.

What error rate per modular addition can we tolerate? The modular
exponentiation consists of $4 L$ modular multiplications each of which
consists of $L$ modular additions. So for an overall error rate
$\epsilon$ we are allowed an error rate of about $\epsilon/(4
L^2)$. With the random number assumption this says that for comparison
we should look at the $2+2 \log_2 (L) -\log_2 (\epsilon)$ most
significant bits of the two numbers. For definiteness I will choose
$\epsilon=0.01$. So e.g. for $L=1000$ bit numbers we only have to
compare some $2+2\cdot 10+7=29$ bits, thus only a small fraction of
$L$. For estimates of the cost (time and space) of the algorithm I
will leave this small contribution away (note also that the
contribution is not in leading order of $L$).

How many Toffoli gates do we use? First we need 1 Toffoli gate per binary place and then 2 Toffolis for uncomputing the carries. So conditional modular addition cost us $3 L$ Toffoli gates.

Now let's look at the algorithm where we first add up all summands and
only then compute the modulus (mod $N$). The problem is that computing
the modulus of the total sum is not a 1 to 1 function. I propose the
following algorithm: Along with the total sum $s$ we compute an
``approximate total sum'' $s'$ which carries only some of the most
significant bits. To compute $s'$ of course we also only add up the
most significant bits of the summands, thus this doesn't cost us
much. Now we can determine from $s'$ how often we have to subtract
$N$ from $s$ to get the modulus. Finally we run part of this algorithm
backwards and in particular uncompute $s'$. In more detail:

\begin{equation}
0 \to s,s' \to s,s', \lfloor s'/N \rfloor \to s ~mod~ N, s',
\lfloor s'/N \rfloor
\to s ~mod~ N, s' \to s ~mod~ N 
\end{equation}

Here $\lfloor s'/N \rfloor$ means the integral part of $s'/N$. Thus we
compute $s ~mod~ N$ as $s-N \lfloor s'/N \rfloor$. How many of the most
significant bits do we have to use for $s'$? We want the probability
of a wrong modular multiplication to be smaller than $\epsilon/(4
L)$. Thus $s'$ should have some $2+7+\log_2 (L)$ correct bits below the
most significant bit of $N$. To get these bits correct we have to use
some $2+7+2 \log_2 (L)$ bits in each addition when computing $s'$. Being
the sum of $L$ numbers of about the size of $N$, $s'$ will be some
$\log_2 (L)$ bits longer than $N$. So all in all $s'$ will have length
$9+3 \log_2 (L)$ and each addition will also use as many bits. I won't
describe detailed quantum circuits for all this, because at any rate
the cost associated with $s'$ is relatively low. (Exercise for the
ambitious reader: why do I go through the trouble of computing $s'$
separately and not just copy some of the most significant bits of $s$
?)

Let's look at the total number of Toffoli gates for such a modular
exponentiation. We now use the simple (non-modular) conditional
addition circuit described above. To compute the sum and uncompute the
carries it needs 3 Toffoli gates per binary place. Then per modular
multiplication we need $L$ such additions, each of length $L$. We need
another modular multiplication to uncompute the old value of the
running product by using $A^{-1} ~mod~ N$. Then to compute the modular
exponentiation we need $2 L$ such steps. This gives a total of $12
L^3$ Toffoli gates.

\subsubsection{$3 L$ qubits are enough}

So far it seems that this algorithm uses some $5 L$ qubits (in leading
order). To see this let's look at a modular multiplication step:

\begin{equation}
x,p \to x,p,A\cdot p ~mod~ N \to x,A\cdot p ~mod~ N
\end{equation}

Now $x$ is a $2 L$ bit number, whereas $p$ and $A\cdot p ~mod~ N$ are
$L$ bit numbers. Another $L$ qubits of workspace is needed to
temporarily store the carry bits for each addition before they are
uncomputed. Thus we get a total of $5 L$ qubits. Now Manny Knill
(private communication) and possibly others have observed that one can
reduce this to $3 L$ qubits. First let's look at the quantum FFT (that
is Fourier transform of the amplitudes of a register). In Shor's
algorithm this QFFT is the last operation before readout. In this case
the QFFT can be simplified (see appendix) by interleaving unitary
operations and measurements of qubits. This QFFT procedure has the
following structure: Hadamard-transform the most significant qubit of
the register and measure it. Then if the observed value is 1, apply
certain phase shifts to all the other qubits of the register. Then
Hadamard transform the second most significant qubit and measure
it. Again, depending on the measured values of the two most
significant qubits, we have to apply certain phase shifts on the
remaining unobserved qubits. So every step consists of Hadamard
transforming the most significant still unobserved qubit, then
measuring it and then applying certain phase shifts on the remaining
unobserved qubits. The values of these phase shifts depend on the so
far measured qubit-values. The measured values give us the binary
representation of the ``output number''. Note that it is in
bit-reversed order, thus we get the least significant bits first. This
bit-inversion is a feature of the FFT algorithm.

The register we apply this procedure to, is the input register $x$. But
most action happens in the other registers. All that happens to the
input register is the following: It starts out initialized to the
$\ket{0}$ state, then we make the uniform amplitude superposition of
all possible input values $x$ by Hadamard transforming each
qubit. Then each of these qubits controls a modular multiplication,
thus it decides whether the multiplication is done or not (in each of
the quantum-parallel computations). And finally the $x$ register
undergoes the QFFT procedure described above. So after having
controlled ``its'' modular multiplication a $x$-qubit does nothing
until the final QFFT.

Usually we would imagine that we would go from least significant
modular multiplications to most significant ones (significance =
significance of associated $x$-control bit). But of course the order
in which we multiply doesn't matter, so we can e.g. turn around the
order. Then the most significant $x$-qubits will first be ready for
the QFFT. So we can interleave controlled modular multiplication steps
and QFFT steps, thus after each controlled modular multiplication we
will get another bit of the (classical) final output.

After having been Hadamard transformed right at the beginning, a
$x$-qubit doesn't do anything until it controls ``its'' modular
multiplication. Moreover the $x$-qubits in the uniform-amplitude
superposition are not entangled. Thus we don't really have to prepare
them all at the beginning of the algorithm, but we prepare the
$x$-qubit only just before ``its'' modular multiplication.

Therefore we have eliminated the $2 L$ qubit long $x$-register and in
leading order need only $3 L$ qubits.

\subsection{parallelizing the standard algorithm}

Let's first think of a classical computation of the modular
exponentiation. Clearly there are several possibilities to parallelize
the algorithm, at least if we are ready to use much more space
(bits). The modular exponentiation consists of many modular
multiplications and each such multiplication consists of many
additions. Let's e.g. look at the addition of $L$ number, each of
length $L$. Instead of having just 1 running sum, adding one summand
after the other to it, we could in parallel sum up equal subsets of
the summands and then add together these partial sums. In the most
extreme case we would group the summands in pairs, then add each pair,
then add pairwise these sums etc. Thus we could make the whole sum in
$\log_2 (L)$ steps instead of $L$. Of course this would require on the
order of $L$ additional $L$-bit registers. In reversible computation
the partial sums would also have to be uncomputed, possibly roughly
doubling the cost. Note that the same ideas can also be applied to the
parallelization of the $L$ modular multiplications.

Rough considerations show that with this kind of parallelization one
reduces the time of the computation about in the same proportion as
one increases the space (qubits):

\begin{equation}
T \sim \frac{1}{S}
\end{equation}

In this paper I consider such a space-time tradeoff as too costly,
assuming that qubits will be very expensive, but just in case, it may
be good to keep in mind this possibility for parallelization. 

I propose a technique for parallelizing the individual addition steps
with a much better space-time tradeoff. The technique is described in
detail in the appendix, here I just give a rough outline of the basic
ideas:

Usually we have to do addition binary place by binary place, as any
lower significance bit can change the value of a higher significance
bit. Now a first observation which we exploit, is that the probability
of such a dependence goes down exponentially with the distance of the
two bits in the binary representation, thus we can use the observation
that some ``algorithmic errors'' are tolerable.

The other idea is to chop the two numbers to be added into blocks of
some fixed length. To add tw corresponding blocks we really should
know the value of the carry bit coming from the preceding block. The
idea is to compute the sum of the blocks for both possible values of
this unknown carry bit. In a second step we then go through all blocks
from low to high significance and for each block determine what the
correct carry bit should have been.

The first step takes time proportional to the length of a block and
the second step proportional to the number of blocks. Thus, by
choosing block-length and the number of blocks about equal, we get a
square root speed up.

\section{Using FFT-based fast multiplication}

The above ``standard algorithm'' can directly handle a modular
multiplication. If we want to use fast integer multiplication
techniques, it seems that we have to compose the modular
multiplication out of several regular multiplications, at least I
haven't found a better way to do it. Thus we use:

\begin{equation}
p \cdot A ~mod~ N = p\cdot A - N \cdot \lfloor p \cdot \frac{A}{N} \rfloor
\end{equation}

Where $\lfloor \rfloor$ means integral part. Note that of course we
precompute $\frac{A}{N}$ classically, and that we only need it to some
$L$ significant digits. Thus we have to compute some three $L$ times
$L$ bit multiplications. I will show later how to do all this reversibly.

2There is a way to speed up the multiplication of large integers by
using the fast Fourier transform technique, where the Fourier
transform is performed over the ring of integer modulo something. By
iterating this technique one obtains the famous Sch\"onhage-Strassen
algorithm \cite{SoS,hop} with complexity of order $O(n \ln n \ln \ln
n)$ to multiply two $n$ bit integers. Although for very large numbers
this is the fastest known algorithm, it is seldomly used because up to
quite large numbers other ways of speeding up multiplication are
faster. In particular the $O(n^{\log_2 3})$ karatsuba-Ofman algorithm
or its variations are used. I won't use it because it seems to use too
much space, as I show in the appendix. On the other hand the FFT based
multiplication technique is naturally parallel without using much more
space.

\subsection{Outline of FFT-multiplication}

Here I give a rough outline of how one can speed up multiplication
using FFT. Things are described in more details in the appendices. 

Say we want to multiply two $L$-bit numbers. First we split each of
them into $b$ blocks of size $l=L/b$. Then the multiplication consists
essentially of convolutions:

\begin{equation}
a \cdot b = \sum_i \left( \sum_j a_j b_{i-j} \right) 2^{l \cdot i} \qquad
\mbox{where} \quad a=\sum_i a_i 2^{l \cdot i} 
\quad b=\sum_i b_i 2^{l \cdot i} 
\end{equation}

where the range of the summation indices has to be figured out
carefully. The expression in parenthesis is known as a convolution.

The Fourier transformation comes in because the Fourier transform of a
convolution of two functions is the pointwise product of the Fourier
transforms of the functions. For the discrete Fourier transform this
is true for the discrete convolutions appearing in the above
equation. The discrete Fourier transform of the numbers $a_n$ is given
by:

\begin{equation} \label{DFT}
\tilde a_n = \frac{1}{\sqrt{N}} \sum_{m=0}^{N-1} a_m \omega^{n \cdot m} 
\qquad \mbox{where} \quad \omega = e^{2 \pi i /N}
\end{equation}

Using the fast Fourier transform algorithm (FFT) one can now compute
the convolutions according to the scheme:

\begin{equation}
a \cdot b = FFT^{-1} [ FFT[a] \cdot FFT[b]]
\end{equation}
 
where the multiplication on the right hand side means pointwise
multiplication. Pointwise multiplication is of course much easier to
compute than the convolution and furthermore it can trivially be done
in parallel. This procedure can help save time because of the great
efficiency of the FFT. A problem is that we really want to make exact
integer arithmetic, but with the usual FFT we use real (actually
complex) numbers and can compute only to some accuracy. Still this
technique is sometimes used \cite{knuth} and the accuracy is chosen
high enough so that rounding at the end always gives the correct
integers.

For higher efficiency one generalizes the Fourier transform to the
ring of integers modulo some fixed modulus. Apart from the
$1/\sqrt{N}$ term, the discrete Fourier transform can be defined over
any ring with any $\omega$. The point is that we still want to be
able to use the FFT algorithm and of course still want the convolution
theorem to be true.

For the fast Fourier transform algorithm to work, we need $\omega^N
=1$. For the convolution theorem still to be true we furthermore need
the condition:

\begin{equation}
\sum_{j=0}^{N-1} \omega^{j p} =0 \qquad \mbox{for all} \quad 0 < p < N 
\end{equation}

This is e.g. true in the ring mod 13 for $\omega=6$ and $N=12$, thus
in particular $6^{12} ~mod~ 13=1$. Because we also have to compute the
inverse Fourier transform, we must furthermore demand that $\omega$
has an inverse in the ring, but this is usually no problem. We will
compute the FFT over a modulo ring. The convolution theorem is then
modified in that we will get the modulus of the intended
results. Usually we will therefore simply choose the modulus larger
than any possible result, so that taking the modulus doesn't change
anything. Sometimes we will also use the Chinese remainder
theorem. Thus we will compute the convolution with respect to 2
relatively prime moduli which are individually too small, but which
allow us to recover the correct result.

The FFT algorithm is most efficient for $N$ a power of 2. Furthermore
for their algorithm Sch\"onhage-Strassen chose $\omega=2$ or some small power of 2, and the
modulus a power of 2 plus 1. This makes the operations in the FFT very
easy because multiplication with $\omega^n$ is just a shift in binary
and also the modulus is easy to compute. More precisely,
Sch\"onhage-Strassen chose the modulus $\omega^{N/2}+1$ for which the
above conditions are always fulfilled. I will also adopt this choice. 

To get the cost (space and time) of the algorithm we need to know the
``parameters'' $b$ and $\tilde l$. $b$ is the number of blocks and
$\tilde l$ is essentially the block length rounded to the next power
of 2. The ring over which we compute the FFT will be the ring of
integer modulo $(2^{2 \tilde l}+1)$, thus we will handle numbers with
$2 \tilde l$ bits. Both numbers $b$ and $\tilde l$ are of the same
order of magnitude, namely either $b=4*\tilde l$ or $b=2*\tilde
l$. The product $b \cdot \tilde l$ is usually $2 L$ rounded to the
next power of 2. Thus

\begin{equation}
b \cdot 2 \tilde l = 4 L \dots 8 L
\end{equation}

depending on $L$. Again, things are described in detail in the appendix.

\subsection{2-level FFT-multiplication}

As mentioned above the Sch\"onhage-Strassen algorithm iterates the FFT
multiplication technique. Thus the component-wise multiplication of
$FFT[a]$ and $FFT[b]$ is again done with FFT-multiply and so on. I
propose to use FFT-multiply on two levels.

I have investigated various algorithms, like 1-level FFT-multiply with
parallelized addition for the component-wise multiplications. For
simplicity I will only present the algorithms which have turned out to
perform well. This is in particular 2-level FFT-multiply with the
individual operations in the FFT parallelized, as described in the
appendix.

An important point is that the $2^{nd}$ level FFT-multiply is much
more efficient than the $1^{st}$ level (for the same size of
numbers). Remember that the component-wise multiplications are actually
done modulo some modulus of the form $m=2^n+1$. It turns out that this
can be done directly with some minor modification of the
FFT-multiply. This modification consists of some multiplication with a
square root of $\omega$. To make this simple we have to choose
$\omega$ an even power of 2. 

As for the first level FFT we have two parameters that characterize
the second level FFT, namely $b'$ and $\tilde l'$. Where $b'$ is the
number of blocks and $\tilde l'$ is the number of bits per block. The
modulus relative to which we calculate the FFT is $2^{2 b'}+1$. 


\subsection{computational cost of 2-level FFT-multiply}

First let's see how to compute the modular multiplication
reversibly. I propose the straight forward scheme:

\begin{equation}
p \to p, \lfloor p \cdot \frac{A}{N} \rfloor \to
p, \lfloor p \cdot \frac{A}{N} \rfloor,
N \lfloor p \cdot \frac{A}{N} \rfloor \to
p,N \lfloor p \cdot \frac{A}{N} \rfloor \to
p,p A - N \lfloor p \cdot \frac{A}{N} \rfloor 
\end{equation}

In every one of these 4 steps we have to compute a multiplication and
then uncompute the garbage. This is of the form:

\begin{equation}
p \to G(p,A), p A \to p, p A
\end{equation}

Thus for a modular multiplication we need a total of 8 simple
multiplications of the form $p \to G(p,A), p A$. For a 1-level
FFT-multiply this would look as follows:

\begin{equation}
p \to \tilde p \to \tilde p, \tilde p \cdot \tilde A \to 
\tilde p, \widetilde{\tilde p \cdot \tilde A} \to 
\underbrace{\tilde p, \widetilde{\tilde p \cdot \tilde A}}_{=G(p,A)}, p A
\end{equation}

Where the tilde stands for Fourier transform and $\tilde p \cdot
\tilde A$ is a component-wise product. 

For a 2-level FFT-multiply we have:

\begin{equation}
p \to \tilde p \to G'(\tilde p,\tilde A), \tilde p \cdot \tilde A \to 
G'(\tilde p,\tilde A), \widetilde{\tilde p \cdot \tilde A}, p A
\end{equation}

where $G'$ represents the lower level garbage which is produced when
$\tilde p$ and $\tilde A$ are multiplied component-wise using the
second level FFT-multiply. So one multiplication consists of 2 first
level FFT's and $b$ lower level multiplications. So for the modular
multiplication (MM) we can write:

\begin{equation}
1 MM =8 (2 FFT_1 + b \times 1 m)
\end{equation}

Where $1 m$ stands for 1 lower level multiplication. Schematically
the step $\tilde p \to G'(\tilde p,\tilde A), \tilde p \cdot \tilde A$
is done as follows:

\begin{equation} \label{mspace}
(\tilde p = b \times p') \to 
b \times (\widetilde{p'}, \widetilde{p'} \cdot \widetilde{A'}, 
\mbox{carries}) \to
b \times (\widetilde{p'}, 
\widetilde{\widetilde{p'} \cdot \widetilde{A'}}, p' A') =
G'(\tilde p,\tilde A), \tilde p \cdot \tilde A 
\end{equation}

Where the carries are used as work space during the lowest level
multiplication which is done as in the standard algorithm. One lower
level multiplication consists of 2 lower level FFT's and $b'$ lowest
level multiplications (denoted by $\mu$). Thus now:

\begin{equation} \label{MM}
1 MM =8 (2~ FFT_1 + b \times (2~ FFT_2 + b' \times 1 \mu))
\end{equation}

\subsubsection{space requirements}

We have to look where in the algorithm most qubits are used. Because
FFT-multiply is relatively space-intensive, this is during the lower
level FFT-multiply and in particular when during the lowest level
multiplication the carries are used for addition
(eq. \ref{mspace}). In this step $\tilde p'$, $\tilde p' \cdot \tilde
A'$ and the carries each use space $b' \cdot 2 b'$ (as the modulus is
$2^{2 b'}+1$). In the modular multiplication scheme we see that apart
from this, there is also sometimes an $L$-bit number around, but for
simplicity I neglect this relatively small contribution. Thus all in
all we get:

\begin{equation}
S=b \times 3 (b' \cdot 2 b')=b \times 3 \cdot 2 (2 \tilde l \dots 4 \tilde l)
= 12 b \tilde l \dots 24 b \tilde l = 24 L \dots 96 L
\end{equation}

where I have used that $b' \cdot b' = 2 \tilde l \dots 4 \tilde l$ and
$b \tilde l = 2 L \dots 4 L$, depending on the size of $L$ relative to
the next power of 2. On average (averaging over $\log L$) the space
requirements are about $50 L$, thus closer to $24 L$ than to $96 L$.

\subsubsection{time requirements}

For this we first have to know the cost of the FFT which consists of
steps of the form:

\begin{equation}
a,b ~\to~ (a+b) ~mod~ (2^n+1), (a-b) 2^m ~mod~ (2^n+1)
\end{equation}

where $a,b$ are non-negative and smaller than $2^n+1$. How to do this
and what it costs is investigated in the appendix. There we get for
the number of Toffoli gates for such an operation on two $n$-bit
numbers:

\begin{equation}
T=13 n
\end{equation}

where $T$ can be thought of as standing for either ``time'' or ``Toffoli''.

Now without further measures the first level FFT would dominate the
time of the parallelized 2-level algorithm. So in the appendix I also
show how to parallelize the individual FFT-operations. The number of
Toffoli gates per individual FFT-operation then rises to:

\begin{equation}
T=26 (n+14)
\end{equation}

but because of the parallelization, the execution time becomes
approximately a constant (for our range of input numbers):

\begin{equation}
T_p=540
\end{equation}

where $T_p$ means the execution time of the parallelized algorithm,
measured in units of the time needed for one Toffoli gate. 

Plugging all this into equation (\ref{MM}) we get the number of
Toffoli gates for 1 modular multiplication. Multiplying this with $4 L$
gives the number of Toffoli gates for the whole modular exponentiation:

\begin{equation}
T =4 L \cdot 8 \left(2~ \log_2(b) \cdot \frac{b}{2} \cdot 26 (2 \tilde l+14) + 
b \times (2~ \log_2(b') \cdot \frac{b'}{2}\cdot 13 (2 \tilde b')+
b' \times 3 (2 \tilde b')^2) \right)
\end{equation}

Remember that for the first level FFT we use the parallel algorithm
whereas for the second level algorithm we use the standard
algorithm. In an FFT the basic operation is carried out $\log_2(b)
\cdot b/2$ times. For the standard modular multiplication which is
used at the lowest level, there is a 3 for the cost of a conditional
addition.

Let's now obtain the execution time of the parallelized algorithm. For
this we have to drop the factors $b/2$, $b'/2$ and $b'$, and plug in
the execution time for the parallelized basic FFT-operation:

\begin{equation}
T_p =4 L \cdot 8 \left( 2~ \log_2(b) \cdot 540 + 
\left(2~ \log_2(b') \cdot 13 (2 \tilde b') +
3 (2 \tilde b')^2 \right) \right)
\end{equation}

\section{Results}

The dominant part of Shor's quantum factoring algorithm is modular
exponentiation. I have found 2 (reversible) algorithms for modular
exponentiation which perform well for factoring very large numbers. In
particular I have introduced much parallelism while still using $O(L)$
space (qubits).  I have looked at numbers of up to several million
digits and some of the approximate results below are only valid for
this range.

Also I have given an improved version of the standard version of
Shor's algorithm (which uses $O(L^3)$ gates, where $L$ is the number of
bits of the number to be factored). The first of the 2 fast algorithms
is a version of this standard algorithm with a parallelized addition.
It's performance is summarized in the following 3 quantities:
\begin{eqnarray}
S &=& 5 L \\
T &=& 52 L^3 \\
T_p &=& 600 L^2
\end{eqnarray}

Where $S$ is the number of qubits used in the algorithm, $T$ is the
total number of Toffoli gates and $T_p$ is the execution time of the
parallelized algorithm measured in execution times of a single Toffoli
gate. Note that the result for $T_p$ is approximate and only valid for
the range of $L$ we are looking at.

For very large numbers an FFT-based multiplication algorithm is used
to compute the modular multiplications in the modular
exponentiation. The FFT-multiply technique is iterated once, thus it
is a 2-level FFT-multiply. The main virtue of this algorithm is that
it is naturally parallel without using much more space (qubits). The
performance of this algorithm depends on the value of $L$ relative to
the next power of 2 and it is not easy to give closed form expressions
for $S$, $T$ and $T_p$. The expressions for $T$ and $T_p$ come from
rough fits to the graph below.

\begin{eqnarray}
S & = & 24 L \dots 96 L \\
T & \approx & 2^{17} L^2 \\
T_p &\approx & 2^{17} L^{1.2}
\end{eqnarray}

$T$ and $T_p$ are plotted as the thick solid lines in the following
graph. The 3 thin solid lines refer to the standard algorithm with
parallelized addition. The two straight lines are $T$ and $T_p$ for
this algorithm. For a fair comparison with the FFT-based algorithm we
have to give this algorithm the same amount of space, even though the
space-time tradeoff of the additional parallelization achievable with
this additional space is not good. This is what the zig-zagged thin
solid line represents. I have assumed that the extra space can be used
for parallelization with space-time tradeoff of the form $T_p \sim
1/S$. The zig-zag is because the amount of space used by the FFT-based
algorithm has such a non continuous behavior.

The straight dotted line is the standard algorithm with $S=3 L$ and
$T=12 L^3$. 

Note that the graph is logarithmic in the number $L$ of bits of the
number to be factored and also in $T,T_p$.

\resizebox{15cm}{13cm}{\includegraphics[120,250][550,566]{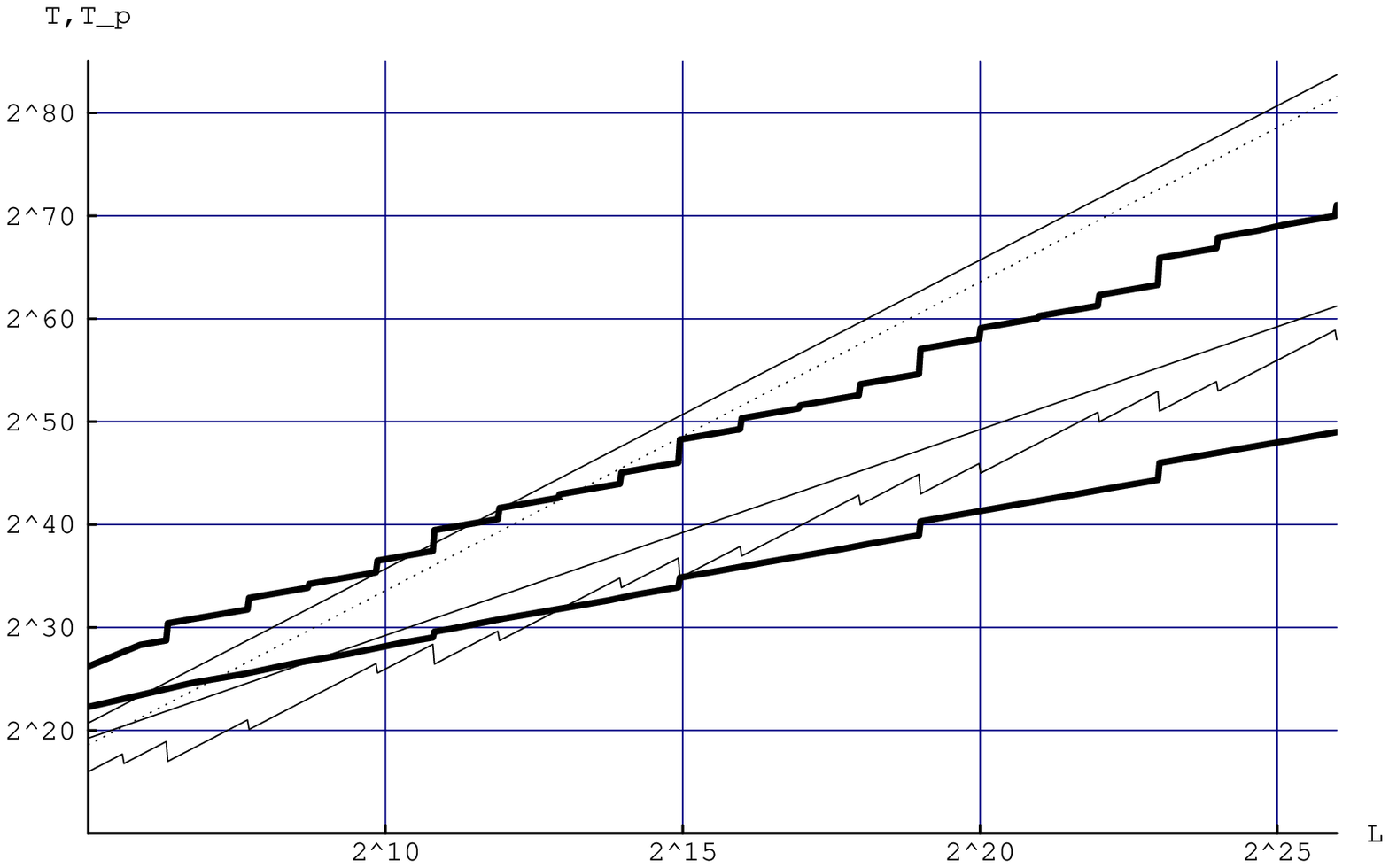}} \\

\subsection{discussion}

The thin zig-zag line and the lower thick line are of most interest as
they give the execution times of the two algorithms (given the same
amount of space for both). We see that they cross around $L \approx
2^{13} \approx 8000$. Thus to factor numbers of more than 8000 bits
the FFT-based algorithm is preferable. (This is provided that we have
the necessary $S=24 L \dots 96 L$ qubits available and not e.g. only
$S=3 L$.)

Given a quantum computer with space and speed comparable to todays
conventional PC's but fully parallel and well ``connected'', the
FFT-based algorithm could be used to factor numbers with maybe up to
millions of binary digits. Consider e.g. a number with $L=2^{20}
\approx 10^6$ binary digits. We would need something around $S=50$
Million qubits\footnote{\underline{computational} qubits, the actual
number of physical qubits will be higher by some factor depending on
the fault tolerant scheme used} ($\approx$ 6 Million
qubytes). Assuming a Toffoli execution-time of $1 \mu s$ the
computation would take around 1 month. The standard algorithm would
take some $2^{24}$ times more time.

Note that the computational resources of the universe are only enough
to factor numbers with several thousand decimal digits when using the
best known classical factoring algorithms.

The possibility that very large numbers could be factored once
sizable quantum computers can be built, may be of interest already
now, as it shows that very large RSA-keys would have to be used to
ensure that a message remains unreadable for several decades to come.

\section{Appendices}

\subsection{parallelizing addition}

The following is a known technique in classical computation: To add
two $L$ bit numbers, decompose them each into $b$ blocks each of
length $l$, thus $L=b \cdot l$. We can't simply add the corresponding
blocks in parallel to get the sum because we don't know the carry bit
coming from the preceding block. The idea now is to compute for each
block (really each pair of blocks) both possibilities, once assuming
that the carry bit coming from the preceding block is 0 and once that
it is 1. This takes time $O(l)$. Then starting with the least
significant block, we go through all blocks, each time determining what
the correct carry for the next block is and from that which of the two
trial-additions was correct and what the correct carry is for the next
block. This takes time $O(b)$. This method can also be iterated, thus
e.g. a scheme with $L=b' \cdot b \cdot l$ will take time $O(b'+b+l)$.

Because we are allowed to make ``algorithmic'' errors for a small
fraction of the input values, we can speed up things even further. The
probability that flipping a block carry bit will change such a carry
bit much further up (in the number) is small for generic summands,
namely 2 to the minus the number of bits between them. Thus the
$2^{nd}$ step of the above algorithm can be somewhat parallelized. Say
we group the blocks into $b'$ superblocks each containing $b''$
blocks, thus $b=b' \cdot b''$. Now we first assume that the input
carry bit to each superblock is 0 and compute the sequence of correct
block carry bits within each superblock. The probability that the
outgoing carry bit of the superblock is wrong because we used 0 as an
incoming carry bit is small. Of course this is done in parallel for
all superblocks. So now that we have the (most probably correct)
superblock input carry bits we compute a second sequence of
block-carry bits for each superblock.

Now let's look at a reversible version of this. Remember that we want to add a fixed (``classical'') number to a quantum register. I propose the
following scheme:

\begin{eqnarray}
a &\to& g_0 \to g_0,g_1 \to g_0,g_1,p \to g_0,g_1,p,f \to g_0,g_1,p,f,f1,f2 \\
  &\to& g_0,g_1,p,f,f1,f2,a+A \to a,a+A
\end{eqnarray}

This may need some explanation. $a$ is the (``quantum-'') number to
which we want to add the fixed number $A$. Actually the above sequence
of operations doesn't manage to get rid of the input $a$ directly, so
it will have to be called twice to uncompute $a$ is a $2^{nd}$
step. $g_0$ and $g_1$ are the 2 numbers we get by once guessing 0 as
input carry for all blocks and once 1. These 2 numbers also include
the output carries of all blocks. $g_0$ can be computed from $a$
without leaving the input $a$ around. Then we compute $g_1$ from
$g_0$. Because we leave the input around, such an addition can
actually be done without carry bits and therefore also without the
need to uncompute them.

$p$ denotes the provisional block carry bits which are determined by
assuming that the input carry to each superblock is 0. Then $f$ is the
final such carry series. Thus when we want to assemble the final sum,
$f$ tells us for each block whether we should take it from $g_0$ or
$g_1$. 

Remember that we want to do the addition depending on a conditional
qubit $e$ ($e$ for ``enable''). For this purpose I use $f_1$ and
$f_2$. $f_1$ is the bitwise AND of $f$ with $e$ and $f_2$ is the
bitwise AND of $\bar f$ and $e$. We have a quantum register
initialized to $\ket{0}$ ready for the final sum $a+A$. We now copy
$g_0$ into it if for the block the corresponding bit of $f_1$ is 1 and
then we copy $g_1$ into it if the corresponding bit of $f_2$ is 1. By
``copy'' I really mean XOR. Because this XOR depends on $f_i$, we
actually need a Toffoli gate for ``copying'' each binary place.

After that we uncompute all the intermediate quantities to get back
$a$. Thus we are left with $a$ and if $e$ is 1, with
$a+A$. Schematically:

\begin{equation}
a, e~ \mbox{AND}~ (a+A) = a, e~ \mbox{AND}~ s
\end{equation}

In a second run of the sequence we bitwise XOR the first register with
the second one minus $A$. Again we do this only if $e=1$. Thus
depending on the enable bit, we are left with either $0,s$ (when
$e=1$) or $a,0$ (when $e=0$). In order to have the result for both
cases in the same register, we have to swap (exchange) the bits in the
two registers depending on $e$. To do this I propose to first XOR the
$1^{st}$ register into the $2^{nd}$ which gives $0,s$
resp. $a,a$. Then , depending on $\bar e$ we XOR the $2^{nd}$ into the
$1^{st}$, which costs us a Toffoli gate per binary place.

Now let's look how this algorithm performs. How much space do we need?
To get $g_0$ and $g_1$ we need $2 (L+b)$ qubits. This is when we store
the carry bits for getting $g_0$ in the space for $g_1$, which forces
us to compute $g_0$ and $g_1$ one after the other. Then we need
another $4 b$ qubits for the 4 versions of block carry bits and
finally another $L$ for the final sum. Thus $S=3 L+6 b$. Below we will
see that for the range of $L$ in which we are interested, $b\approx
L/6$ and thus $S \approx 4L$. The standard addition costs $2 L$, thus
parallelization here costs about $2 L$ additional qubits.

Let's count the total number of Toffoli gates: Computing $g_0$ costs
$2 L$ Toffolis as it is an unconditional addition. Computing $g_1$
costs only $L$ Toffolis as here we don't even have to uncompute
carries. The bits of $p$ are computed by copying either one of two
bits into it, which one depending on yet another bit. This can
certainly be done with 2 Toffolis. Thus $p$ and $f$ cost us each $2 b$
Toffolis. $f_1$ and $f_2$ cost us each $b$ Toffolis and the copying of
$g_0$ resp. $g_1$ to get the final sum costs another $2 L$
Toffolis. This totals to $5 L+6 b$. But to uncompute $a$ we have to do
all this twice, plus at the end we need $L$ Toffolis to swap the two
registers depending on $e$. Thus a total addition step costs $T=11
L+12 b \approx 13 L$ Toffolis. This compares to only $3 L$ Toffolis for the
standard addition, but we hope that we still get a substantial speed
up due to parallelization.

So how much time does the algorithm take? I again measure this in
(sequential) Toffoli gates. The result may depend more than other
quantities on the unknown details of a quantum computer architecture,
so the following is essentially an educated guess. $g_0$ and $g_1$ are
computed one after the other, but in parallel for all blocks. This
costs some $2 l+l$ Toffoli time steps. The computation of $p$ takes
some $b''$ steps, each involving 2 Toffolis. The same is true for
$f$. It looks like $f_1$ and $f_2$ could be computed in full
parallelism. In practise it may not be possible to use $e$
simultaneously as control bit for many Toffoli gates, thus we may want
to make some copies of $e$ (and maybe also of $\bar e$). Somewhat
arbitrarily I set the time cost to $2 b''$ for computing both $f_i$'s. A
similar situation occurs at the end when we obtain the final sum by
copying from $g_0$ and $g_1$. With the conservative assumption that we
do this sequentially for each block, we get $2 l$ time steps. The
total is $5 l +6 b''$. 

Again this has to be doubled to account for the uncomputation of $a$
and we also have to take into account the conditional swapping of the
registers at the end. For this conditional swapping I assume that
there are enough copys of $e$ to do this in $l$ time steps. Thus a
total addition step costs $T_p=11 l+12 b''$ Toffoli time steps. (Here
the index $p$ stands for ``parallel''.)

Let's now consider how we make the decomposition $L=b' b'' l$ into
superblocks and blocks. The superblocks have to be large enough to
make it very improbable that a carry ``runs all the way through
them''. The error probability per superblock has to be smaller than
the error probability per full addition. Let's first conservatively
set it to $\epsilon/(4 L^3)$. Thus with the usual uniform random
number assumption we get that a superblock should be at least some
$9+3 \log_2 (L)$ bits long. 

Let's not overdo formal generality, and let's assume that $L$ is in
the following range: $L=2^9 \dots 2^{25}$. Then $\log_2$ of the length of a
superblock is about 5 to 6 and we get that the overall error is still
smaller than $\epsilon$ when we reduce the length of a superblock to
about $4+3 \log_2 (L)$.

To chose a reasonable block size within a superblock, note that the
number of blocks per superblock $b''$ and the length of a block $l$
contribute about equally to the computation time $T_p$. So to minimize
$T_p$ we should chose them about equal. So we get the following
approximate Formula for the computation time per addition: $T_p
\approx 23 \sqrt{3+3 \log_2 (L)}$ . For the range $L=2^9 \dots 2^{25}$
we get $T_p \approx 130 \dots 200$. As I will propose to use this algorithm for not too large values of $L$, I simply set $T_p \approx 150$.

Also consider the following table:
\\ 

\begin{tabular}{|r|c|c|c|c|r|r|r|} \hline

$L$ & $3+3 \log_2 L ( = b'' \cdot l)$ & b'' & $l$ & 
$T_p (=  11 l+12 b'')$ & $ 3 L$ & speed up \\ \hline

  500 & 30 & 5 & 6 & 127 &   1500 &   12 \\ \hline

 5000 & 40 & 6 & 7 & 150 &  15000 &  100 \\ \hline

50000 & 50 & 7 & 7 & 161 & 150000 &  900 \\ \hline

\end{tabular} \\ \\ \\

With $T_p = 150$ per addition we get the following comparison of space
and time for the full modular exponentiation:

\begin{eqnarray}
\mbox{conventional:} && \quad S = 3 L \quad T=12 L^3 \\
\mbox{with parallelized addition:} && \quad S = 5 L \quad T=52 L^3 
\quad T_p = 600 L^2
\end{eqnarray}

\subsection{Why not the $O(n^{\log_2 3})$ Karatsuba-Ofman algorithm?} 

In particular there is the Karatsuba-Ofman algorithm with running time
$O(n^{\log_2 3}) \approx O(n^{1.58})$. Actually there is a whole
series of algorithms with ever smaller exponents, which are actually
approaching 1, but here I will only consider the basic case. 

In conventional computation it seems that one of these algorithms
usually beats Sch\"onhage-Strassen for the size of numbers of
interest. Of course for practical applications the computer
architecture, word size etc. also play a big role. It seems that for
reversible computation and assumptions I make about the possible
architecture of a quantum computer, things look differently and a
variant of the Sch\"onhage-Strassen algorithm may actual find an
application.

The reason for this is that I assume that qubits will be very
expensive. Also I look for algorithms which can be massively
parallelized, but again without increasing the work space too much. As
I will show below, Sch\"onhage-Strassen does this nicely, massive
parallelization will not increase the space demand
dramatically. Karatsuba-Ofman can also naturally be parallelized, but,
as I show, it seems that it will use a lot of space for this. The
algorithm is simple, to multiply to $n$-bit numbers we split each of
the numbers in two $n/2$-bit pieces, assuming that $n$ is even. Thus:

\begin{equation}
a \cdot b = (a'+2^{n/2} a'') (b'+2^{n/2} b'') 
= a' b' + (a' b''+a'' b') 2^{n/2} + a'' b'' 2^n
\end{equation}

So we could now compute the product by computing 4 products of
$n/2$-bit numbers, which takes about the same time. But there is the following
simple trick which allows to do it in only 3 such multiplications:

\begin{equation}
(a' b''+a'' b') = (a'+a'') (b'+ b'') -a' b' - a'' b''
\end{equation}

Note that we anyways have to compute $a' b'$ and $a'' b''$. By
iterating this technique, thus applying it again to the smaller
multiplications, we get the asymptotic $O(n^{\log_2 3})$
performance. Clearly the smaller multiplications can be done in
parallel.

Now let's look at the space requirements of such a parallelization. At
the first level we have to store the six $n/2$-bit numbers
$a',b',a'',b'',(a'+a'')$ and $(b'+ b'')$. For each lower level the
space needed is 3/2 times the space needed at the uper level. By
summing up this geometric series and noting that there are some
$\log_2 n$ levels, we get the result that we need total space
$O(n^{\log_2 3})$. This is too much if, as I expect, qubits will be
very expensive. 

Actually even without parallelization the Karatsuba-Ofman algorithm
uses much space. For this we look at the point at which on all levels
the sums $a'+a''$ and $b'+b''$ have been prepared.

\subsection{The fast Fourier transform (FFT) and FFT-multiply}

\subsubsection{the FFT-algorithm}

The discrete Fourier transform of $N$ numbers over the complex numbers 
is:

\begin{equation}
\widetilde{a_n} = \frac{1}{\sqrt{N}} \sum_{m=0}^{N-1} e^{2 \pi i \frac{n \cdot m}{N}} ~a_m
\end{equation}

It can be generalized to numbers out of any ring by replacing $e^{2
\pi i/N}$ by some fixed ring-element $\omega$. In general of course also the
normalization $1/\sqrt{N}$ must be left away:

\begin{equation}
\widetilde{a_n} =\sum_{m=0}^{N-1} \omega^{n \cdot m} ~a_m
\end{equation}

If $\omega^N=1$ and $N$ can be decomposed into small prime factors,
the fast Fourier transform (FFT) -algorithm can be used, which is
widely used on computers. Usually $N$ is a power of 2: $N=2^l$, as
then the FFT is most efficient and best suited for binary digital
computers. I demonstrate the FFT for this case and will also include
the normalization $1/\sqrt{N}$, but it can easily be left away for
rings where there is no $1/\sqrt{2}$.

In the first step of the FFT we reduce the original task to 2 Fourier
transforms over $N/2$ numbers each. This is then iterated $\log_2 N
=l$ times until we are left with (trivial) Fourier transforms of
single numbers.

We start with: 

\begin{equation} 
\widetilde{a_n} =
\frac{1}{\sqrt{N}}\sum_{m=0}^{N-1} \omega^{n \cdot m} ~a_m
\end{equation}

Consider the binary representations of $n$ and $m$:

\begin{equation} 
n=(\overbrace{\underbrace{n_{l-1}, \dots n_2}_{n''=\lfloor n/4 \rfloor}
, n_1}^{n'=\lfloor n/2 \rfloor}, n_0) \qquad \quad
m=(m_{l-1}, \overbrace{m_{l-2}, \underbrace{m_{l-3}, \dots m_0}_
{m'' = m ~mod~ 2^{l-2}}}^{m'= m ~mod~ 2^{l-1}})
\end{equation}

Where the definitions of $n'',m''$ and so on should be clear. The
symbol $\lfloor ~\rfloor$ means the integer part. We now consider
separately the 2 cases $n_0=0,1$ and sum over the 2 values of
$m_{l-1}$:

\begin{eqnarray}
\widetilde{a}_{(n',n_0)} &=& \frac{1}{\sqrt{N/2}} ~\sum_{m'=0}^{N/2-1}
\frac{1}{\sqrt{2}} ~\sum_{m_{l-1}=0}^1 
\omega^{(2 n'+n_0) (2^{l-1} m_{l-1}+m')}
~a_{(m_{l-1},m')} \\
&=& \frac{1}{\sqrt{N/2}} ~\sum_{m'=0}^{N/2-1}
\frac{1}{\sqrt{2}} ~\sum_{m_{l-1}=0}^1 
\omega^{2 n' m'}~ \omega^{n_0 m'} \omega^{2^{l-1} n_0 m_{l-1}}
~a_{(m_{l-1},m')} \\
&=& \frac{1}{\sqrt{N/2}} ~\sum_{m'=0}^{N/2-1} (\omega^2)^{n' m'}~ 
\underbrace{\omega^{n_0 m'} 
~\frac{a_{(0,m')} + (-1)^{n_0} a_{(1,m')}}{\sqrt{2}}}_{{a'}_{(n_0,m')}}
\end{eqnarray}

On the second level we then have:

\begin{equation} \label{bo1}
a''_{(n_0,n_1,m'')} = (\omega^2)^{n_1 m''} 
~\frac{a'_{(n_0,0,m'')} + (-1)^{n_1} ~a'_{(n_0,1,m'')}}{\sqrt{2}}
\end{equation}

and so on. Note that the leftmost bits in the parenthesis are still
the most significant ones. In the end we have:

\begin{equation}
\widetilde{a}_{(n_{l-1}, n_{l-2}, \dots n_0)} = 
a^{(l)}_{(n_0,n_1, \dots n_{l-1})}
\end{equation}

So in the end we have to interpret the bits in the reversed order to
get the numbers of the result in the right order. 

The actual operations we have to do are transformations of the
following form of 2 numbers:

\begin{equation} \label{bo2}
a, ~b \quad \to \quad \frac{a+b}{\sqrt{2}}, ~ \omega^k \frac{a-b}{\sqrt{2}}
\end{equation}

\subsubsection{the quantum fast Fourier transform (QFFT)}

Note that the Fourier transform over the complex numbers is a unitary
transformation of $N$ complex numbers. It is therefore in principle
possible to apply it physically to the $2^l$ amplitudes of an
$l$-qubit quantum register:

\begin{equation}
\ket{\mbox{quantum-register}} = \sum_{n=0}^{2^l-1} a_n ~\ket{n} 
\quad \to \quad \sum_{n=0}^{2^l-1} \widetilde{a_n} ~\ket{n} 
\end{equation}

where the $\ket{n}$ are the ``computational'' basis states given by
the binary representation of $n$. Note that such a transformation is
very different from what is usually done on conventional computers
where the values represented by binary words are transformed. It is
therefore misleading to say that quantum computers are much faster
than conventional computers at the FFT.

The QFFT can be done very efficiently. The FFT is done level by level
($l$ levels) as the operations on every level are individually
unitary. In the classical FFT on every level $N/2$ basic operations of
the form (eq. \ref{bo2}) are carried out. In the QFFT all these $N/2$
operations are done in parallel.

So on level number $l-l'$ we should transform the amplitudes as follows
(see eq. \ref{bo1}):

\begin{equation}
a_{(n_{l-1}, \dots n_{l'}, \dots n_0)} \to 
e^{2 \pi i ~\frac{n_{l'} \cdot (n_{l'-1}, \dots n_0)}{2^{l'+1}}}
~\frac{a_{(n_{l-1}, \dots 0, \dots n_0)} + (-1)^{n_{l'}}
a_{(n_{l-1}, \dots 1, \dots n_0)}}
{\sqrt{2}}
\end{equation}

where the parenthesis mean the value in binary given by the bits.
This operation can be done in 2 steps, first without the phase factor
$e^{i \dots}$ and then multiplying the basis states with $n_{l'}=1$
with appropriate phase factors:

\begin{eqnarray}
a_{(n_{l-1}, \dots n_{l'}, \dots n_0)} & \to &
~\frac{a_{(n_{l-1}, \dots 0, \dots n_0)} + (-1)^{n_{l'}}
a_{(n_{l-1}, \dots 1, \dots n_0)}}
{\sqrt{2}} \\
a_{(n_{l-1}, \dots 1, \dots n_0)} & \to &
e^{2 \pi i ~\frac{(n_{l'-1}, \dots n_0)}{2^{l'+1}}}
~a_{(n_{l-1}, \dots 1, \dots n_0)} 
\end{eqnarray}

The first step is simply a Hadamard transformation on qubit number
$l'$. The conditional rephasing can be decomposed into conditional
rephasings on individual qubits:

\begin{equation}
a_{(n_{l-1}, \dots 1, \dots n_0)} \to
e^{2 \pi i ~\frac{n_{l'-1}}{4}}
~e^{2 \pi i ~\frac{n_{l'-2}}{8}} \dots
e^{2 \pi i ~\frac{n_0}{2^{l'+1}}}
~a_{(n_{l-1}, \dots 1, \dots n_0)} 
\end{equation}

So these are $l'$ gates, each one a phase shift on some qubit,
conditional on qubit number $l'$. Note that in the end we should also
swap bit number $l-1$ with bit number $0$, bit number $l-2$ with bit
number $1$ and so on to get the originally intended result, but
usually it is not necessary to do this explicitly.

Let's summarize in words how the QFFT is carried out: We begin by
Hadamard transforming the most significant qubit. Then we apply
appropriate phase shifts to all less significant qubits, conditional
on the most significant one being 1. Then we Hadamard transform the
second most significant qubit and conditional on it we apply
appropriate phase shifts to all less significant qubits and so on. 

After Hadamard transforming a qubit we don't apply gates to it any
more. Thus if, as in Shor's algorithm, the register is to be observed
right after the QFFT, we can always measure qubits after they were
Hadamard transformed, thus interleaving unitary gates and
measurements.

A further advantage of this is that now the phase shifts are no more
conditional: we do them if the last measured qubit was 1 and don't do
anything otherwise.

Note that most phase shifts are very small, so it is enough to carry
out only the $O(\log_2 l)$ largest one on every level, without
significantly changing the state of the QC.

Also we can wait with applying phase shifts to a qubit until just
before it is Hadamard transformed (and then measured). The size of the
phase shift will then depend on the already measured values of the
higher significance qubits. So now what we do it to apply some phase
shift to a qubit, Hadamard transform and measure it. Then we move on
to the next lower significance qubit.

In fault tolerant quantum computing the phase shift gates will have to
be composed of several gates from a fixed ``set of universal
gates''. Still the QFFT is a negligible part of the factoring
algorithm.

\subsubsection{computing convolutions with the FFT}

Say we have two sets of numbers $a_n$ and $b_n$ with $n=0 \dots
N-1$. Then the convolution is given by:

\begin{equation}
c_n = \sum_{k=0}^{N-1} \sum_{l=0}^{N-1} \delta_{k+l,n} ~a_k ~b_l
\qquad \mbox{where} \quad n=0 \dots 2 N-2
\end{equation}

This can efficiently be computed using the FFT. The statement is
essentially that the component-wise product of the FFT's of the $a$'s
and the $b$'s is the FFT of the convolution. So we try:

\begin{eqnarray} \label{confusion}
c'_n &=& \sum_{m=0}^{N-1} \omega^{-n \cdot m} 
~\widetilde{a}_m ~\widetilde{b}_m \\
&=& \sum_{m=0}^{N-1} \omega^{-n \cdot m} 
\sum_{k=0}^{N-1} \omega^{m \cdot k} a_k
\sum_{l=0}^{N-1} \omega^{m \cdot l} b_l \\
&=& \sum_{k,l} \sum_m \omega^{m (k+l-n)} a_k b_l
= N \sum_{k,l} (\delta_{k+l-n,0} + \delta_{k+l-n,N}) a_k b_l
\end{eqnarray}

Where for the last step I have used:

\begin{equation} \label{cond2}
\sum_{n=0}^{N-1} \omega^{p \cdot n} =0, 
\quad \mbox{except for} \quad p= \mbox{integer} \cdot N
\end{equation}

This is true for $\omega = e^{2 \pi i/N}$ (sum up the geometric
series), but for a general ring it has to be required in addition to
$\omega^N=1$. Note that $\omega^{-1}= \omega^{N-1}$.

Also besides the convolution we have gotten a second unwanted term. It
can be made to vanish by setting the ``upper half'' of the sets $a_n$
and $b_n$ equal to zero. So to compute the convolution of two
$N$-number sets we then have to use FFT's with $2 N$ numbers.

\subsubsection{1-level FFT-multiply}

I assume that the reader has looked at the outline of FFT-multiply in
the main body of the paper. We want to do the FFT over the ring of
integers modulo some fixed element $\omega$. First note that things
will work out for the particular class of moduli $M= \omega^{N/2} +1$,
where $N$ is a power of 2. It is easy to see that, as required,
$\omega^N =1$, where we now do all operations modulo $M$. The other
property (eq. \ref{cond2}) is demonstrated in the book by Hopcroft
et. al. \cite{hop}. We will choose $\omega$ a small power of 2 which
makes the operations in the FFT very efficient on a binary digital
computer.

As mentioned, the numbers to be multiplied are first decomposed into
blocks of bits. The question now is what size of blocks we should
choose. If there were an adequate modulus, we would like to choose the
block size much smaller than the number of blocks. Unfortunately the
modulus proposed above is quite large and so there is no point in
choosing small blocks as in the course of the FFT the numbers we
operate with will quickly become the size of the modulus.

If we insist on a small modulus we must be ready to give up the ease
with which we can operate with the above $\omega$ and $M$. Also it is
not trivial to find such moduli where there are primitive roots with
$\omega^{2^l}=1$ and the condition of eq. \ref{cond2}. It may still
pay off, but I have given up this path as it is rather complicated.

So back to the above choice of $\omega$ and $M$. So how do we choose
the number $b$ of blocks and their size $l$ ? Note that $b$ is the
number of numbers we have to Fourier transform, thus it must be a
power of 2. Then the modulus will be $M=\omega^{b/2} +1$. For the
result not to get truncated the block size has to be somewhat less
than $(\log_2 M)/2$, which means that $b$ and $l$ are both going to be
of the order of magnitude of $\sqrt{L}$, where $L$ is the number of
bits of the numbers we want to multiply.

Without further justification I show how to obtain the $b$ and $l$
which I have found to be optimal. First we set

\begin{equation}
k= \lceil log_2 (2 L) \rceil 
\end{equation}

Then there are 2 cases:

\begin{eqnarray}
k ~\mbox{even:} && b=2^{\frac{k}{2}+1} \quad 
\mbox{and} \quad \tilde l=2^{\frac{k}{2}-1} \\
k ~\mbox{odd:} && b=2^{\frac{k+1}{2}} \quad 
\mbox{and} \quad \tilde l=2^{\frac{k-1}{2}}
\end{eqnarray}

Where $\tilde l$ is the next power of 2 after $l$. And $l$ is:

\begin{equation}
l= \lceil 2 L / b \rceil 
\end{equation}

The modulus is $M=2^{2 \tilde l}+1$. Thus for $k$ even we have
$\omega=2$ and for $k$ odd $\omega=4$. The components of the
convolution are the sum of some $b$ products of numbers with $l$
bits. For the convolution not to be truncated by the modulus we thus
must require:

\begin{equation}
2 l + \log_2 b < 2 \tilde l
\end{equation}

If this condition is not true I simply revise my choice of $b$,
effectively going to the next larger size of the algorithm. Thus in
practise I increase $k$ by 1 and get the new $b$ and $\tilde l$ from
it. There would be more efficient methods, but as this occurs seldomly
for large $L$, I choose the easier way.

\subsubsection{the 2-level FFT-multiply}

One can iterate the FFT-multiply and apply it to the component product
of the Fourier transformed factors. Actually FFT-multiply is much more
efficient on the second level (or still lower levels). On the second
level we have to multiply two $2 \tilde l$ bit numbers modulo $2^{2
\tilde l}+1$ which can be done quite efficiently by modifying the way
we computed the convolution. This time we don't have to pad
the factors with zeros, so the decomposition into blocks is $b' \tilde
l' =2 \tilde l$, where the prime refers to the second level. Note that
now the block size $\tilde l'$ is automatically a power of 2. So a
factor $a$ is decomposed into its blocks $a_n$ as follows:

\begin{equation}
a = \sum_{n=0}^{b'-1} a_n 2^{n \cdot \tilde l'}
\end{equation}

If we now use FFT's for $b'$ numbers to compute the convolution of two
numbers $a$ and $b$, we get the previously unwanted second term (from
eq. \ref{confusion}):

\begin{equation}
c'_n = b' \sum_{k,l} (\delta_{k+l-n,0} + \delta_{k+l-n,N}) a_k b_l
\end{equation}

The final result of the FFT-multiply then becomes:

\begin{equation}
c = \sum_n 2^{n \cdot \tilde l'} 
\sum_{k,l} (\delta_{k+l-n,0} + \delta_{k+l-n,N}) a_k b_l
\end{equation}

The actual product of $a$ and $b$ is

\begin{equation}
c = \sum_n 2^{n \cdot \tilde l'} 
\sum_{k,l} \delta_{k+l-n,0} a_k b_l
\end{equation}

modulo $M=2^{2 \tilde l} +1$ it becomes very similar (same up to a
sign) to what we got using FFT-multiply:

\begin{equation}
c ~mod~ (2^{2 \tilde l}+1)= \sum_n 2^{n \cdot \tilde l'} 
\sum_{k,l} (\delta_{k+l-n,0} - \delta_{k+l-n,b'}) a_k b_l
\end{equation}

FFT-multiply can actually be modified to get just that change of
sign. For this we need a square root $\psi$ of $\omega'$. Then consider

\begin{eqnarray}
c'_n &=& \sum_m (\omega')^{-n \cdot m} 
\sum_k (\omega')^{m \cdot k} ~\psi^k a_k
\sum_l (\omega')^{m \cdot l} ~\psi^l b_l \\
&=& \sum_k \sum_l b' (\delta_{k+l,n}+\delta_{k+l,n+b'}) \psi^{k+l} a_k b_l \\
&=& \psi^n \sum_k \sum_l b' (\delta_{k+l,n}-\delta_{k+l,n+b'}) a_k b_l
\end{eqnarray}

Thus by doing $a_n \to \psi^n a_n$ and $b_n \to \psi^n b_n$ we get
$c_n \psi^n$. We will choose $\omega'$ a small even power of 2 so that
$\psi$ is a small power of 2. These operations (which are always done
modulo $M'$) are rather easy. How to do them reversibly is described at
the end of the appendix about how to compute reversibly the FFT. From
there it should be clear that their cost can be neglected.

I will chose the modulus $M'=2^{2 b'}+1$ so $\omega'=16$ (and thus
$\psi=4$). Again we have to consider separately two cases (where
$k'=\log_2(2 \tilde l)$):

\begin{eqnarray}
k' ~\mbox{even:} && b'=2^{\frac{k'}{2}} \quad 
\mbox{and} \quad \tilde l'=2^{\frac{k'}{2}} \\
k' ~\mbox{odd:} && b'=2^{\frac{k'+1}{2}} \quad 
\mbox{and} \quad \tilde l'=2^{\frac{k'-1}{2}}
\end{eqnarray}

For $k'$ odd the modulus $M'$ is safely large, so there won't be any
problem with the result getting truncated by computing the remainder
$~mod~ M'$.

For $k'$ even the modulus is not large enough. We can fix this problem
by also making an FFT-multiply with the modulus $2^{b'}+1$ and using
the Chinese remainder theorem to recover the correct result. The
computational cost for this should be negligible. Here I just give a
short outline without derivations. So say we are looking for a
non-negative number $x$ which is smaller than $(2^{2 n}+1)(2^n+1)$ and
we are given $x'=x ~mod~ (2^n+1)$ and $x''=x ~mod~ (2^{2 n}+1)$. Then:

\begin{equation}
x = x'' + (2^{2 n}+1) \left( (2^{n-1} (x'-x'')) ~mod~ (2^n+1) \right)
\end{equation}

where I have used that

\begin{equation}
( - (2^{2 n}+1)) ~mod~ (2^n+1) = 2^{n-1}
\end{equation}

To see how to compute $x$ in reversible computation without leaving
much ``garbage'' qubits around, see again the end of the appendix
about how to compute the FFT reversibly.

\subsection{How to compute the FFT in reversible computation}

\subsubsection{introduction}

Here I show how to carry out the basic operations of the FFT. In
principle every one of these steps is reversible and it thus seems
that the FFT-algorithm is particularly well suited for reversible
computation\footnote{please remember that this FFT has nothing to do
with the QFFT which Fourier transforms amplitudes}. Here we consider
the FFT over the ring of integers modulo a modulus of the form
$M=2^n+1$ and with the primitive root $\omega$ a small power of 2. The
basic FFT-operations acting on 2 registers are then of the form:

\begin{equation} \label{FFTop}
a,b \quad \to \quad (a+b) ~mod~ (2^n+1), ~(a-b) 2^m ~mod~ 
\underbrace{(2^n+1)}_{=M}
\end{equation}

where $a,b$ are non-negative and smaller than $2^n+1$. It is easy to
see how this operation could be reversed, as 2 has a multiplicative
inverse modulo any odd modulus. Thus we could use the standard tricks
(eqs. \ref{rev1}, \ref{rev2}) of reversible computation to get rid of
garbage and input, but usually one can find more efficient
``shortcuts'' which is what we will try here. Actually I haven't found
an efficient way of doing it without leaving anything but the result
(RHS of eq. \ref{FFTop})) around, so I propose a scheme which leaves 2
or 3 qubits of garbage around. This is no problem as it doesn't take
up much space, but I would have expected to find a more elegant
solution. So everybody interested is invited to try to do better. Note
that any FFT we perform will soon afterwards be undone, so then the
garbage qubits will also go away.

The way to attack the problem is to decompose it into several steps
each of which is in itself reversible. We have the following 3 steps:

\begin{eqnarray}
\mbox{1. step:} \quad && a,~b \quad \to \quad 
a,~b+a \quad \to \quad 2 a-(b+a), ~b+a \\
\mbox{2. step:} \quad && a+b,a-b \quad \to \quad 
(a+b) ~mod~ M, ~(a-b) ~mod~ M \\
\mbox{3. step:} \quad && (a-b) ~mod~ M \quad \to \quad (a-b) 2^m ~mod~ M
\end{eqnarray}

The 1. step consists of additions of 2 ``quantum'' numbers, which is
no big problem. The 2. step consists of conditionally subtracting $M$
from $a+b$ and conditionally adding $M$ to $a-b$. The problem here is
how to uncompute the conditional bits. The 3. step looks rather simple
as this operation is easy in conventional computation, but here I
haven't managed to find an efficient scheme without leaving garbage
around.

\subsubsection{``quantum''+ ``quantum'' addition: $a,b \to a, b+a$}

The 1. step above consists of 2 such operations. A subtraction is
essentially the same using the complement technique. One would expect
that an addition of a ``quantum'' number to another ``quantum'' number
would use more Toffoli gates than the previously described addition of
a fixed ``classical'' number to a quantum number, but this is not
so.\footnote{I admit that after deciding to only count Toffoli gates I
was tempted to minimize their number, which may not be quite clean}
Note that one of the ``quantum'' numbers has to stay around as
otherwise the operation would not be reversible.

So lets do $a,b \to a, b+a$. Thus we add $a$ to the
$b$-register. Again we temporarily need carry qubits (which are
initially set to 0). I denote by $c_i$ the carry which comes from the
$(i-1)^{st}$ binary place but has the same place value as $a_i$ and
$b_i$. For every binary place there are 2 operations, one to compute
the next carry $c_i$ from $a_i,b_i,$ and $c_i$ and one to compute the
sum bit:

\begin{equation}
c_{i+1} = (a_i+b_i+c_i \ge 2) \qquad \quad b_i \to a_i \oplus b_i \oplus c_i
\end{equation}

where the parenthesis means a logical expression and $\oplus$ is
addition modulo 2 (or XOR). For some reason I prefer to first
calculate the sum bit. Then we have:

\begin{equation}
b_i \to a_i \oplus b_i \oplus c_i \qquad \quad 
c_{i+1} = (a_i+\bar{b_i}+c_i \ge 2) 
\end{equation}

The following sequence of quantum gates accomplishes this:

\begin{picture}(300,100)

\thinlines \multiput(70,0)(0,25){4}{\line(1,0){250}} \thicklines
\put(0,75){$a_i$} \put(0,50){$b_i$}
\put(0,25){$c_i$} \put(0,0){$c_{i+1}$}

\put(90,0){
\put(0,50){\line(0,1){25}} \put(0,75){\C} \put(0,50){\NOT}}

\put(110,0){
\put(0,25){\line(0,1){25}} \put(0,25){\C} \put(0,50){\NOT}}

\put(180,0){\put(0,50){\NOT}}

\put(200,0){
\put(0,50){\line(0,1){25}} \put(0,75){\C} \put(0,50){\NOT}}

\put(220,0){
\put(0,25){\line(0,1){50}} \put(0,75){\C} \put(0,25){\NOT}}

\put(240,0){\line(0,1){50} \put(0,50){\C} \put(0,25){\C} \put(0,0){\NOT}}

\put(260,0){
\put(0,50){\line(0,1){25}} \put(0,75){\C} \put(0,50){\NOT}}

\put(280,0){
\put(0,0){\line(0,1){75}} \put(0,75){\C} \put(0,0){\NOT}}

\put(300,0){\put(0,50){\NOT}}

\end{picture} \\

To check the correctness of this sequence and of the above formulas I
recommend considering separately the case $a_i=0$ and $a_i=1$. The
first 2 gates (from the left) compute the sum and the rest is for the
carry. After having computed the sum we must run everything backwards
to uncompute the carries. But because we don't want to uncompute the
sum, we then leave away the 2 leftmost gates.

\subsubsection{computing the modulus}

Here we want to do:

\begin{equation}
a+b,~a-b \quad \to \quad \underbrace{(a+b) ~mod~ M}_{= \Sigma = a+b-f_+ M}, 
~\underbrace{(a-b) ~mod~ M}_{= \Delta = a-b+f_- M} \\
\end{equation}

Note that $f_+$ and $f_-$ are either 0 or 1. We will have to compute
these 2 conditional bits:

\begin{equation}
f_+ = (a+b \ge M) \qquad f_- = (a-b <0)
\end{equation}

Actually we begin by subtracting $M$ from $a+b$ and add it again if we have
obtained a negative number, but of course $f_+$ will remain around. As
we use complement notation for negative numbers, $f_-$ is trivial to
obtain. Then we make a conditional addition to get $a-b+f_- M$. 

So now we have the result and would like to get rid of the conditional
bits. To do this is equivalent to computing those bits from the result
$\Sigma, \Delta$. As $M$ is odd it is easy to obtain the XOR of the
two bits. This can be seen as follows:

\begin{equation}
\Sigma + \Delta =a+b -f_+ M +a-b + f_- M = 2 a +(f_- - f_+) M
\end{equation}

Thus if this sum is even we know that $f_+ = f_-$ and vice versa. This
we read off simply from the lowest significance bits of $\Sigma$ and
$\Delta$. So now we can reduce the garbage to 1 qubit. To get rid of
this remaining bit seems to be more costly and I propose not to do
it. Nevertheless I show how it could be done. We have to consider two
cases:

\begin{eqnarray*}
f_+ = f_- = f : && \qquad \Delta - \Sigma = 2 f M - 2b \quad \mbox{so} \quad
f=(\Delta > \Sigma) \\
1- f_+ = f_- = f : && \Sigma + \Delta = 2 a +(1- 2 f) M \quad \mbox{so} \quad
f=(\Sigma + \Delta \ge M)
\end{eqnarray*}

Note that because we are doing this in quantum parallelism we have to
compute always both expressions. This is quite costly and thus I prefer
to leave a garbage qubit around.

\subsubsection{computing ~$x \cdot 2^m ~mod~ (2^n+1)$}

Here $x=(a-b) ~mod~ (2^n+1)$ and let $y=x 2^m$. For this
modulus we can relatively easily compute the remainder of any $y$ by
decomposing it into blocks $y_i$ of $n$ bits:

\begin{equation}
y ~mod~ (2^n+1) = \sum y_i 2^{i \cdot n} ~mod~ (2^n+1) 
= \sum y_i (-1)^i ~mod~ (2^n+1)
\end{equation}

For our $y$ we have only 2 terms in the sum. The potentially non-zero
bits in these two blocks overlap in exactly 1 bit, as $y$ has $n+1$
potential non-zero bits. When we compute the sum $\sum x_i 2^{i \cdot
n}$ we will have to leave around one of these 2 bits. One of those
bits is the $(n+1)^{st}$ bit of $x$ and this is the one I propose to
leave around. After this we still may have to add $2^n+1$ if the sum
is negative to get the correct remainder. I also propose to leave
around the associated  control qubit as I haven't found an easy way to uncompute
it.

\subsubsection{assessment of total cost}

\begin{eqnarray}
a,b \to a, a+b && \qquad 2 n \\
a,a+b \to 2 a- (a+b), a+b && \qquad 2 n \\
a+b \to (a+b \ge M), (a+b) ~mod~ M && \qquad 3 n \\
a-b \to (a-b < 0), (a-b) ~mod~ M && \qquad 3 n \\
\Sigma+ \Delta \stackrel{?}{=} \mbox{odd} && \qquad O(1) \\
(a-b) ~mod~ M \to (a-b) 2^m ~mod~ M, ~2~ \mbox{garbage qubits} && \qquad 3 n
\end{eqnarray}

The first 2 lines are (unconditional) ``q+q'' additions. Uncomputing
the carries doubles the number of Toffoli gates from $n$ to $2 n$. The
next 2 lines are essentially conditional addition of the ``c+q'' type,
the same we use in the standard algorithm. The cost of the last line
comes from the conditional addition of $M$. So the total number of
Toffoli gates per elementary FFT-operation is:

\begin{equation}
T= 13 n
\end{equation}

For an FFT with $b$ numbers this has to be multiplied by $ \log_2(b)
\cdot b/2$, as there are $\log_2(b)$ levels and on each level $b/2$
elementary operations are carried out.

\subsubsection{parallelizing this}

For large $n$ we want to parallelize the additions in the above
scheme, furthermore for large $n$ we can make substantial
simplifications which will only lead to a small ``algorithmic'' error
rate. Without such improvements the first level FFT in my proposed
2-level FFT-multiply scheme would dominate the overall execution
time. I have estimated that for the range of $L$ we consider, an error
rate per elementary FFT-operation of about $2^{-40}$ still leads to a
tolerable overall error rate.\footnote{exercise for the ambitious
reader} As on the first level FFT $n$ will be larger than 40 we can
make simplifications based on the special form of the modulus
$M=2^n+1$. First we can assume that all numbers modulo $2^n+1$ are
actually smaller than $2^n$.
\footnote{this is due to the uniformly distributed random number
assumptions, which, I admit, I'm not very confident about in this
case} Note that this will also reduce the garbage from 3 to 2 bits, as
we can assume that the other one is zero (it's the $(n+1)^{st}$ bit of
$x$).

In the above list of costs of the individual operations the three
lines with cost $3 n$ can be simplified as they essentially involve an
addition or subtraction of $M=2^n+1$. Note that the condition bit $a+b
\ge M$ is replaced by $a+b \ge 2^n$ which is trivial to compute. Also
adding or subtracting $2^n$ is simple. Adding or subtracting 1 costs a
bit more. To keep the error rate low enough, we need to extend the addition
to the 40 least significant bits. Because it is a conditional addition
this costs $3 \cdot 40$ Toffoli gates. Also the second and third lines
can actually be done simultaneously.

What remains are the first two lines which are ``q+q'' type
additions. It turns out that my scheme for parallelizing ``c+q'' type
additions works just as well for ``q+q'' additions. From there we get
$T=13 n$ and $T_p=150$ per addition, where $T$ is the total number of
Toffoli gates and $T_p$ is the number of sequential Toffoli gates,
thus essentially the time measured in units of Toffoli execution
times.

Taking all this together we get for the parallelized and simplified
version of the elementary FFT-operation:

\begin{eqnarray}
T &=& 2 \cdot 13 n + 3 \cdot 40 \cdot 3 \approx 26 (n+14) \\
T_p &=& 2 \cdot 150 + 2 \cdot 40 \cdot 3 = 540
\end{eqnarray}

\end{document}